\documentclass{emulateapj}
\submitted{{\it Submitted for publication in ApJ}}
\usepackage{multirow,color,wrapfig}
\usepackage {graphicx}
\usepackage{graphics}
\usepackage[dvips]{epsfig}

\def\be{\begin{equation}}
\def\ee{\end{equation}}
\def\ba{\begin{eqnarray}}
\def\ea{\end{eqnarray}}

\newcommand{\beq}{\begin{eqnarray}}  
\newcommand{\eeq}{\end{eqnarray}}

\newcommand{\ly}{{\ifmmode{{\rm Ly}\alpha}\else{Ly$\alpha$}\fi}}
\newcommand{\hMpc}{{\ifmmode{h^{-1}{\rm Mpc}}\else{$h^{-1}$Mpc}\fi}}  
\newcommand{\hGpc}{{\ifmmode{h^{-1}{\rm Gpc}}\else{$h^{-1}$Gpc}\fi}}  
\newcommand{\hmpc}{{\ifmmode{h^{-1}{\rm Mpc}}\else{$h^{-1}$Mpc}\fi}}  
\newcommand{\hkpc}{{\ifmmode{h^{-1}{\rm kpc}}\else{$h^{-1}$kpc}\fi}}  
\newcommand{\hMsun}{{\ifmmode{h^{-1}{\rm {M_{\odot}}}}\else{$h^{-1}{\rm{M_{\odot}}}$}\fi}}  
\newcommand{\Mmin}{{\ifmmode{{M_{\rm min}}}\else{${M_{\rm min}}$}\fi}}
\newcommand{\mmin}{{\ifmmode{{M_{\rm min}}}\else{${M_{\rm min}}$}\fi}}
\newcommand{\mmax}{{\ifmmode{{M_{\rm max}}}\else{${M_{\rm max}}$}\fi}}
\newcommand{\lmmin}{{\ifmmode{{\log M_{\rm min}}}\else{${\log M_{\rm min}}$}\fi}}
\newcommand{\lmmax}{{\ifmmode{{\log M_{\rm max}}}\else{${\log M_{\rm max}}$}\fi}}

\newcommand{\Mmax}{{\ifmmode{{M_{\rm max}}}\else{${M_{\rm max}}$}\fi}}
\newcommand{\dm}{{\ifmmode{{\Delta M}}\else{$\Delta M$}\fi}}
\newcommand{\dlm}{{\ifmmode{{\Delta \log M}}\else{$\Delta \log M$}\fi}}
\newcommand{\focc}{{\ifmmode{{f_{\rm occ}}}\else{${f_{\rm occ}}$}\fi}}

\newcommand{\Msun}{{\ifmmode{{\rm {M_{\odot}}}}\else{${\rm{M_{\odot}}}$}\fi}}  
\newcommand{\msun}{{\ifmmode{{\rm {M_{\odot}}}}\else{${\rm{M_{\odot}}}$}\fi}}

\newcommand{\rand}{{\ifmmode{{\mathcal{R}}}\else{${\mathcal{R}}$ }\fi}}  
\newcommand{\reff}{{\ifmmode{r_{\mbox{\tiny eff}}}\else{$r_{\mbox{\tiny eff}}$}\fi}}
\newcommand{\todo}{\ifmmode \text{\Huge{\(\bullet\)}} \else {\Huge$\bullet$}\fi}
\newcommand{\tido}{\ifmmode {\bullet} \else $\bullet$\fi}

\begin{document}
%=========================================================================
%		FRONT MATTER
%=========================================================================
\title{Impact of cosmic variance on the galaxy-halo connection for Lyman-$\alpha$ emitters }
\author{
  Juli\'an E. Mej\'ia-Restrepo \thanks{jemejia@das.uchile.cl}$^{1,3}$,
  Jaime E. Forero-Romero \thanks{je.forero@uniandes.edu.co}$^{2}$ 
}

\affil{
$^1$Departamento de Astronom\'{i}a, Universidad de Chile, Camino el Observatorio 1515, Santiago, Chile\\
$^2$Departamento de F\'{i}sica, Universidad de los Andes, Cra. 1
No. 18A-10, Edificio Ip, Bogot\'a, Colombia\\
$^3$FACom-Instituto de F\'isica-FCEN, Universidad de Antioquia, Calle 70 No. 52-21, Medell\'in, Colombia
}

\begin{abstract}
 In this paper we study the impact of cosmic variance and observational uncertainties
in constraining the mass and occupation fraction, \focc, of
dark matter halos hosting \ly\ Emitting  Galaxies (LAEs) at high redshift. 
To this end, we construct mock catalogs from an N-body simulation to match the 
typical size of observed fields at $z=3.1$ ($\sim 1 {\rm deg^2}$).
In our model a dark matter halo with mass in the range $\mmin
<M_{\mathrm h}<\mmax$ can only host one detectable LAE at most.    
We proceed to  explore the parameter space determined by \mmin, \mmax\ and \focc\
with a Markov Chain Monte-Carlo algorithm using the angular correlation function (ACF) and the LAEs number density  as observational constraints. 
We find that the preferred minimum and maximum masses in our model span
a wide range $10^{10.0}\hMsun\leq \mmin \leq 10^{11.1}\hMsun$ ,
$10^{11.0}\hMsun\leq \mmax \leq 10^{13.0}\hMsun$; followed by a
wide range in the occupation fraction $0.02\leq \focc \leq 0.30$.   
As a consequence the median mass, $M_{50}$, of all the consistent models has a large uncertainty $M_{50} =
3.16^{+9.34}_{-2.37}\times 10^{10}$\hMsun.
However, we find that the same individual models have a relatively tight
$1\sigma$  scatter around the median mass $\Delta M_{1\sigma} = 0.55^{+0.11}_{-0.31}$ dex.
We are also able to show that \focc\ is uniquely determined by $M_{\rm
  min}$, regardless of $M_{\rm max}$. 
We argue that upcoming large surveys covering at least $25$ deg$^{2}$
should be able to put tighter constraints on \mmin\ and \focc\ through the LAE number density
distribution width constructed over several fields of
$\sim 1$ deg$^{2}$.
\end{abstract}


%*************************************************************************
\section{Introduction}
\label{sec:introduction}

Lyman-$\alpha$ emitting galaxies (LAEs) are central to a wide range
of subjects in extragalactic astronomy. 
LAEs can be used as probes of reionization \citep[for a recent review
  see][and references therein]{Dijkstra14}, tracers  of large scale structure
\citep{Koehler2007}, signposts for low metallicity stellar
populations \citep[for a recent review see][and references
  therein]{Hayes15}, markers of the galaxy formation process at high
redshift \citep{Partridge67,Rhoads00,Blanc11} and tracers of active
star formation.    
 
In most of those cases, capitalizing the observations requires
understanding how LAEs are formed within an explicit cosmological
context.  
Under the current structure formation paradigm, the dominant matter
content of the Universe is dark matter (DM).  
Each galaxy is thought to be hosted by a larger dark matter structure
known as a halo. \citep{Peebles1980,SpringelNature05}.  
Understanding the cosmological context of LAEs thus implies studying
the galaxy-halo connection.  
Galaxy formation models suggest that the physical processes that
regulate the star formation cycle are dependent on halo mass
\citep[e.g.][]{Behroozi2013a}. Therefore, the mass becomes 
the most important element in the halo-galaxy connection.  
   
The goal becomes finding the typical DM halo mass of halos hosting LAEs.
In the case of LAEs there are different ways to find this mass range.
One approach is theoretical, using general astrophysical principles to
find the relationship between halo mass, intrinsic \ly\ luminosities
and observed \ly\ luminosities. 
This approach is usually implemented through semi-analytic models
\citep{Garel2012,Orsi2012} and  full N-body hydrodynamical simulations
\citep{Laursen2007, Dayal2009, ForeroRomero2011, Yajima2012}.  

The downside of these calculations is the uncertainty in the
estimation of the escape fraction of \ly\ photons. 
Given the resonant nature of the \ly\ line, the escape fraction is
sensitive to  the dust contents, density, temperature, topology and
kinematics of the neutral Hydrogen in the interstellar medium (ISM). 
The process of finding a consensus on the expected value for the
\ly\ escape fraction in high redshift galaxies is still matter of open
debate
\citep{Neufeld1991,Verhamme2006,ForeroRomero2012,Dijkstra2012,Laursen2013,Orsi2012,Yajima14}.       

A different approach to infer the typical mass of halos hosting
LAEs is based on the spatial clustering information. 
This approach uses the fact
that in CDM cosmologies the spatial clustering of galaxies on large
scales is entirely dictated by the halo distribution
\citep{Colberg00}, which in turn has a strong dependence on halo
mass. 
Using measurements of the angular correlation function of LAEs,
observers have put constraints on the typical mass and occupation
fraction of the putative halos hosting these galaxies
\citep{Hayashino2004,Gawiser07,Nilsson2007,Ouchi2010,Bielby16}. 
In these studies the observations are done on fields of $\sim 1$ deg$^{2}$ and
the conclusions derived on the halo host mass do not elaborate deeply on the
uncertainty resulting from the cosmic variance on these fields. 
For instance, \citet{Guaita2010} attempted to analytically 
estimate the uncertainty from cosmic variance 
following \citet{Somerville2004} and \citet{Peebles1980}. 
Their prescription assumed that the correlation function can be
represented by a power-law
$\xi\left(r\right)=\left(r_{0}/r\right)^{\gamma}$. 
Their estimation of cosmic variance strongly depends on $r_{0}$ and
$\gamma$.  
However, $\gamma$ is only poorly constrained with current
observations which in turns indicates that their cosmic variance
estimation could not be precise. 

In this paper we investigate the impact of cosmic variance in
constraining the mass and occupation fraction of halos hosting LAEs at $z=3$.
We build mock surveys from a cosmological N-body simulation to compare them
against the observations of \cite{Bielby16} using the angular
correlation function.  
We use a simple model to populate a halo in the simulation with a LAE   
assuming a minimum, \mmin, and maximum mass, \mmax, for the dark
matter halos hosting LAEs.  We do not  assume  an underlying relation 
between the \ly\  luminosity and the dark matter halo mass.  
This approach bypasses all the physical uncertainties associated to
star formation and radiative transfer. 
We then use the Markov Chain Monte Carlo technique to obtain the
likelihood of the parameters given the observational constraints.
This approach allows us to estimate cosmic variance directly from
simulations without making any assumption  on the correlation function
behavior. 

Throughout this paper we assume a $\Lambda$CDM cosmology with the
following values for the cosmological parameters, $\Omega_{m}=0.30711$,
$\Omega_{\Lambda}=0.69289$ and $h=0.70$, corresponding to the matter
density, vacuum density and the Hubble constant in units of 100 km
s$^{-1}$ Mpc$^{-1}$.  The values are consistent with \citet{Planck2014} results.

\begin{figure}
\includegraphics[width=0.47\textwidth]{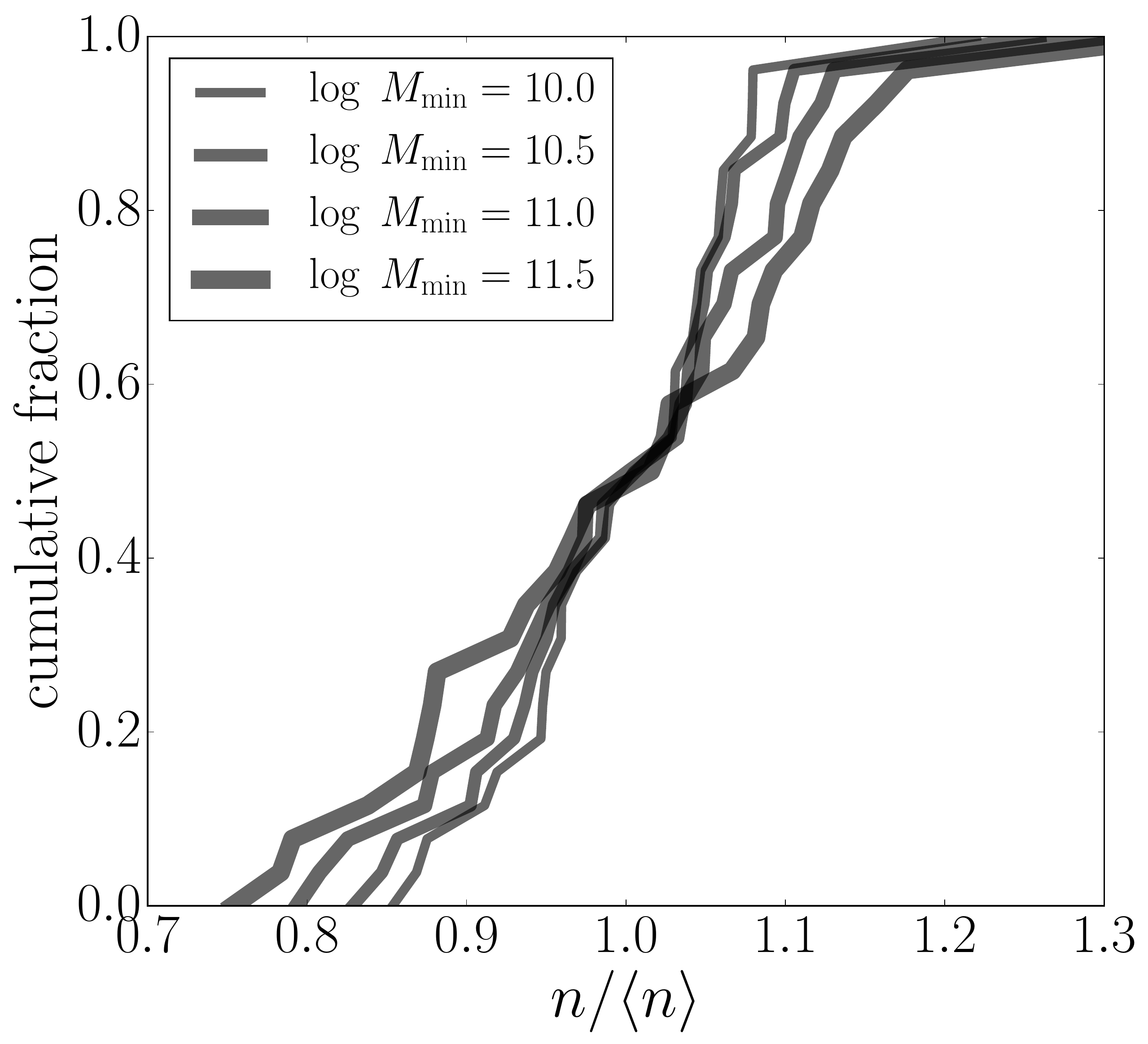}
\caption{Cumulative halo number density distribution function over
  27 mock fields. Each line corresponds to a
  different model with increasing values of \mmin. 
  Different models produce different number density distributions. 
  The width of the distribution increases with \mmin. } 
\label{fig:cosmicv0}
\end{figure}

\section{Methodology}

The base of our method is the comparison between observations and mock
catalogs. 
This approach allows us to take explicitly into account cosmic variance. 
The comparison has four key elements. 
First, the observations we take as a benchmark. 
Second, the N-body simulation and the halo catalogs we use to build
the mocks. 
Third, the parameters describing our model to
assign a LAE to a halo. 
Fourth, the statistical method we adopt to compare observations and
simulations.  
We describe in detail these four elements in the following subsections.

\subsection{Observational constraints}
\label{subsec:obs}
\citet{Bielby16} used narrow band imaging to detect 643 LAE candidates
at $z\sim 3$  with Ly-$\alpha$ rest-frame equivalent widths
$\gtrsim$65\AA\ and a  Ly-$\alpha$ flux limit of $2\times10^{17}{\rm
  erg/cm^2/s}$ ($L\sim 7\times10^{42}{\rm erg/s} $).  
Using follow-up spectroscopy they found a 22\% contamination fraction $f_{\rm c}$.
Their observations cover 5 (out of 9) independent and adjacent
fields of the VLT LBG Redshift Survey (VLRS).  
The  total observed  area corresponds to 1.07$\rm deg^2$ that translates to
$\sim$80$^2\hMpc^2$ in a comoving scale. 
\citet{Bielby16} used the NB497  narrow-band filter whose 77\AA\ FWHM
and 154\AA\ Full width  tenth maximum (FWTM) correspond to a total observational comoving depth of
44\hMpc\  and 82\hMpc, respectively. 

\begin{figure}
  \includegraphics[width=0.47\textwidth]{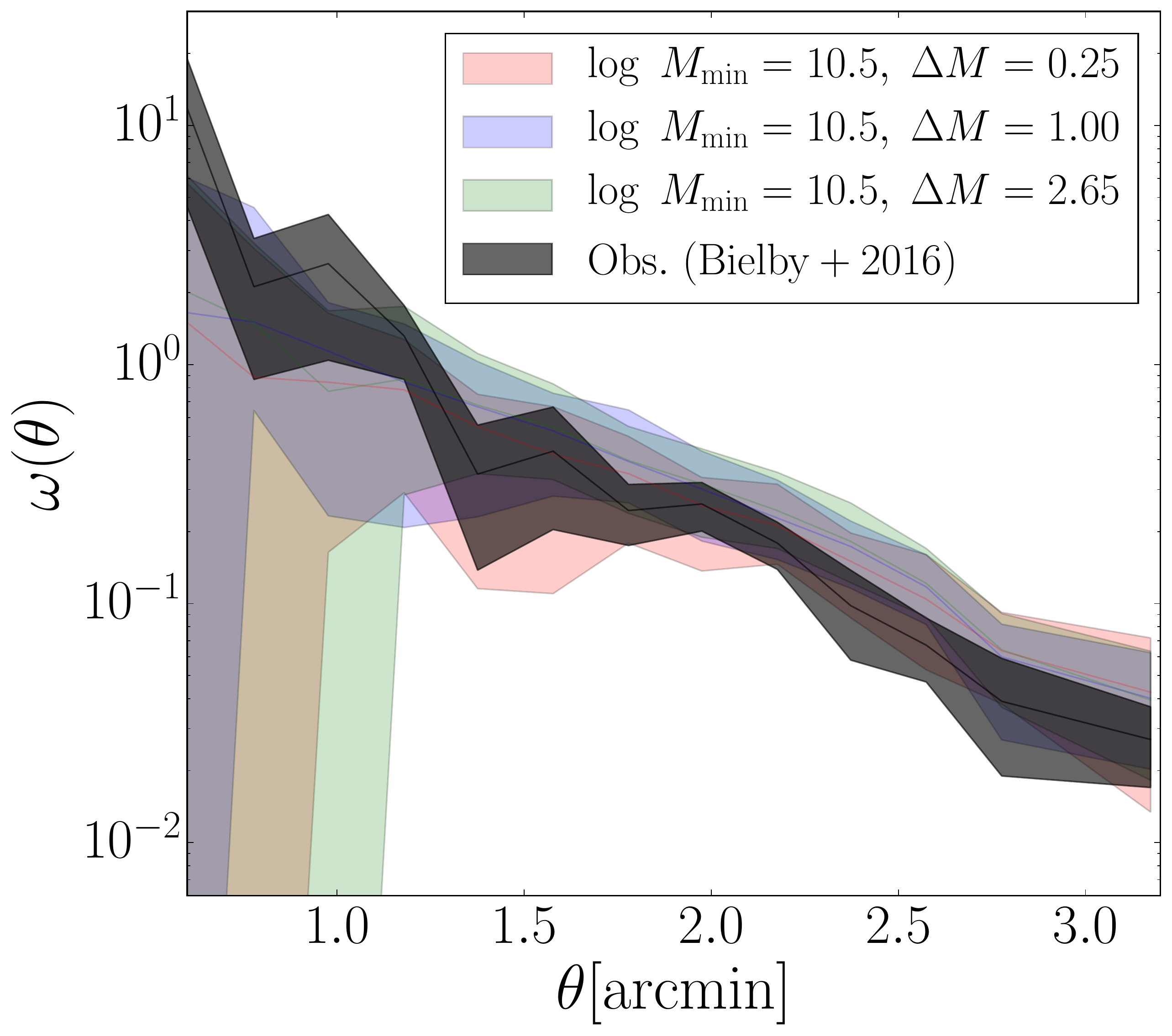}
\caption{ Angular Correlation functions for $\log M_{\rm
    min}[\rm{M_{\odot}h^{-1}}]=10.5$ and different values of $\Delta \log
  M$.  
  The shaded region in the models represents the $1-\sigma$ variation
  due to cosmic variance. Radically different models in $\Delta \log M$ are consistent with
  observations once cosmic variance is modeled in detail.} 
\label{fig:corr}
\end{figure}

\subsection{Simulation and halo catalog}
\label{subsec:sim}

We use results from the BolshoiP simulation \citep{Bolshoi,BolshoiP} 
performed in a cubic  comoving volume of 250\hMpc\  on a side. 
The dark matter distribution was sampled using  $2048^{3}$
particles. 

We chose this simulation based on three requirements. 
First, the simulation has cosmological parameters consistent
with the most recent constraints from Cosmic Microwave Background
experiments and provide public acces to the data.
Second, the halo mass function from the simulation is robust
down to  $10^{10}$\hMsun, which is the lower halo mass bound suggested
by previous observational 
studies \citep{Hayashino2004,Gawiser2007,Ouchi2010,Bielby16}. 
Third, the total volume of the simulation is at least 10 times
larger than the equivalent volume of a single LAE observational field.
These conditions are met as we describe below. 

The cosmological parameters are consistent with Planck
results \citep{Planck2014} with a matter density 
$\Omega_{\rm m} = 0.307$, cosmological constant
$\Omega_{\Lambda}=0.693$, dimensionless Hubble constant $h=0.678$, slope
of the power spectrum  $n=0.96$ and normalization of the power
spectrum $\sigma_{8}=0.823$.  
This translates into a particle mass of  $m_{\rm p}=1.5\times 10^{8}$
$h^{-1}$ M$_{\odot}$. 
Data is available to the
public through an online
interface \footnote{\url{http://www.multidark.org/MultiDark/}}
\citep{MultiDark}.

We use halo catalogs constructed with a Bound-Density-Maxima (BDM)
algorithm. 
For each  halo in the box we extract its comoving position and mass.  
We focus our work on halos more massive than $6.0\times
10^{9}$\hMsun\ resolved with at least  $40$ particles 
to guarantee statistical significance and a well behaved halo mass function. 
We do not take into account sub-halos.

We split the simulation volume at z$\sim$3 into  27 smaller mock
volumes mimicking the  area and depth reported in \citet{Bielby16} and
described in \S \ref{subsec:obs}.

\begin{figure*}
\begin{center}
\includegraphics[width=0.85\textwidth]{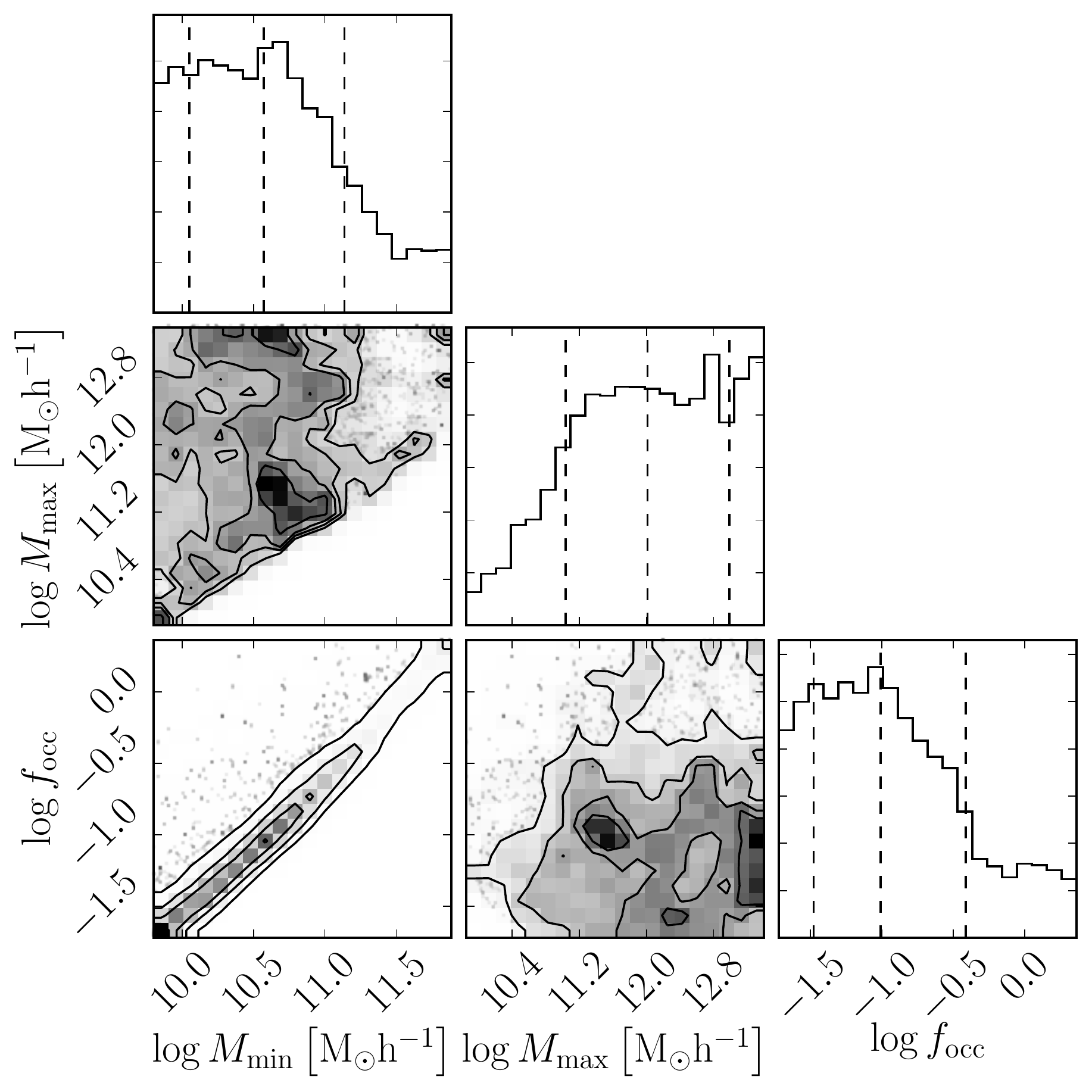}
\end{center}
\caption{One and two dimensional projections of the posterior
  probability distributions of \mmin, \mmax\ and \focc. 
  Vertical lines in the one dimensional proyections correspond to the
  14-50,84 percentiles of the distribtions.
  Thick lines in the two dimensional planes mark the 1-2-3 $\sigma$
  contours of the corresponding joint distributions.
  The models with $\log \focc>0.00$ ($\focc>1$) correspond  to models
  where the number density  of halos is smaller than the number of
  observed LAEs but we consider them as consistent because of the uncertainty
  in the median number density due to cosmic variance. See 
Fig. \ref{fig:cosmicv0} and \S\ref{subsec:explore} for details.} 
\label{fig:like}
\end{figure*}

\subsection{A simple LAE model}
\label{subsec:mocks}

We build the simplest possible model to assign a LAE to each DM halo
without trying to compute a LAE luminosity.
  
We first assume that a dark matter halo can host one detectable LAE at
most.   
This assumption is consistent with theoretical analysis of the
correlation function \citep{Jose2013b} and observations that confirm a
lack of  class pairs in LAEs \citep{Bond2009}.  
Then we say that a halo will host a LAE with probability $\focc$
if and only if the halo mass is in the range $\mmin < M_{\mathrm
  h} < \mmax$. 
We also use the variable \dlm\ to represent the mass range width,
$\dlm\equiv\ \log \mmax - \log \mmax$.

An astrophysical interpretation of $f_{\rm occ}$ convolves at least
four phenomena: the actual presence of a star forming galaxy in a
halo, a duty cycle in the star formation rate, the escape fraction of
\ly\ radiation and its detectability as a LAE with Ly-$\alpha$ equivalent widths of $\gtrsim$65\AA.   
We do not try to disentangle these effects and do not assume an underlying relation between the
$\alpha$ luminosity and the dark matter halo mass.
We instead opt for a purely arithmetic interpretation by setting
$\focc$ as the ratio between the observed number of LAEs and the
number of halos within the considered mass range, that is $\focc \equiv N_{\rm
  LAEs}/N_{\mathrm{halos}}$.   

For each mock catalog we also randomly remove a fraction $f_{\rm c}=0.22$  of the mock-LAEs and replace 
them with randomly distributed points to mimic the effect of interloper
contamination   in \citet{Bielby16} observations. 
On top of that we apply rejection sampling  to our LAE selection 
along the radial direction taking the transmission function of the
NB479 filter used in \citet{Bielby16} observations.

Fig. \ref{fig:cosmicv0}  shows the cumulative halo number density
for all 27 sub-volumes in the simulation, with a normalization by the
median number density among fields, $\langle n \rangle$. 
Each line represents a different model $\mathcal{M}$ with fixed
$\focc=1$ and $\dlm=1.0$; and varying $\mmin$, 
This Figure shows that the halo number  density varies across
sub-volumes, as an expression of cosmic variance. 
As a consequence, the $\focc$ also varies across the mock fields  
by the same factor factor.

In what follows we will describe by the letter ${\mathcal M}$  a model
defined by a particular choice of the two parameters $M_{\rm min}$ and
$M_{\rm  max}$. For each model  ${\mathcal M}$ we define
$\tilde{f}_{\rm occ}$ as the median occupation fraction within the the
mock fields.

\subsection{Model Selection}
\label{subsec:explore}

We  explore the parameter space of the models
${\mathcal M}$ by  means of the Affine Invariant  Markov Chain Monte Carlo technique using
the EMCEE python package \citep[][and references therein]{emcee2013}.

The MCMC exploration is done using a total of 24 seeds and 400
iterations (9600 models) to sample the posterior  probability distribution function, $P(\mathcal{M}|\mathrm{observations})$, based on the Angular
Correlation Function (ACF). We put a flat prior on $\log M_{\rm min}$ and $\log \mmax$ between
$9.8$ up to $13.4$, corresponding to the halo mass range of
the simulation at $z=3$.
We restrict the selection to models that give a minimal number density
$N_{\rm{halos}}>N_{\rm LAE}/3$.
This means that it is possible to have $N_{\rm{halos}} < N_{\rm LAE}$
and hence $\focc\equiv N_{\rm{LAE}}/N_{\rm{halos}}>1$.
We include the $1/3$ factor to account for the uncertainty in the number
density of LAEs due to cosmic variance, expecting it to be on the
same order as the dark matter halo cosmic variance shown in
Fig. \ref{fig:cosmicv0}.   Our likelihood is taken proportional to
$ \exp(-\chi_{\mathcal  M}^2/2)$  where:

\begin{equation}
\chi_{\mathcal M}^2=
\sum_{\theta}\left[\frac{\left( \rm{ACF}_{\mathcal
      M}\left(\theta\right) - \rm{ACF}_{\rm
      obs}\left(\theta\right)\right)^2}{ \sigma_{\rm \mathcal
      M}^{2}\left(\theta\right) + \sigma_{\rm
      obs}^{2}\left(\theta\right)}\right]
\end{equation}
Here  $\rm{ACF}_{\mathcal M}$ and  $\rm{ACF}_{\rm obs}$ are the ACF
of the explored model ${\mathcal M}$ and the observational ACF
reported by \citet{Bielby16}, respectively.
$\sigma_{\rm \mathcal M}$ is the associated 1-$\sigma$ scatter  of the
$\rm{ACF}_{\mathcal M}$ as a product of cosmic variance and
$\sigma_{\rm obs}$ is the observational error associated to
$\rm{ACF}_{\rm obs}$.

We do not include the covariance matrix due to the
practical impossibility to estimate it in a robust way.
The small size of the fields and galaxies make it
unfeasible the application of jackknife and bootstrap techniques \citep{2009MNRAS.396...19N}. 
An estimate from N-body simulation faces the same problem with the
additional limitation that we want to solve in the first place. 
Namely, not knowing the range of halo masses that produces the
observed clustering signal.  
Tests that we performed in the simulation show that, due to the small
number of points to estimate the clustering, the covariance matrix is
not stable, even for similar models. 
Using the full error covariance matrix, when not noise
dominated, versus only diagonal elements usually has a small effect
on clustering analyses \citep{2011ApJ...736...59Z}.

We compute the $\rm{ACF}_{\mathcal M}$ using the Landy \&  Szalay
estimator  \citep{Landy1993}. After this,  we correct the computed ACF
from the contamination fraction of interlopers $f_{\rm c}$ multiplying the
ACF by a factor of $1/(1-f_{\rm c})^{2}$  following the  procedure described in
 \citet{Bielby16}.

Fig. \ref{fig:corr} shows the observational ACF by
\citet{Bielby16} compared to the ACF in three different models with a
wide range in $\dlm$.  
This already shows that radically different models can be compatible
with observations once cosmic variance uncertainties are modeled in
detail.

\section{Results and Discussion}
\label{sec:results}

\subsection{Constraints on Model Parameters}

Fig. \ref{fig:like} shows the one and two dimensional projections of
the posterior probability distributions of the parameters in our LAE model. 
This Figure represents our main result: \mmin, \mmax\  and \focc\ cannot
be tightly constrained from the available observations. 

The preferred $1-\sigma$ range for the masses is $10.0<\log \mmin<11.2$
and $11.0<\log\mmax<13.0$.
\focc\ is completely determined by \mmin\ from \focc$=$0.004 when
$\log\mmin=10.0$ to $\focc=0.38$ when $\log\mmin=11.15$. 
We compute the power-law dependence between \focc\ and \mmin\ to be 
\begin{equation}
\focc = 0.05\left(\frac{\mmin}{10^{10}\hMsun}\right)^{0.77}.
\label{eq:focc}
\end{equation}
We remind the reader that models with $\log \focc>0.00$ ($\focc>1$)
correspond to cases where the halo number density is smaller 
than the number of observed LAEs but are  still considered consistent
because of the expected uncertainty in the median number density of
LAEs due to cosmic variance (see Fig. \ref{fig:cosmicv0} and
\S\ref{subsec:explore}).

The result in Eq.(\ref{eq:focc}) can be qualitatively understood as
follows.   
A choice of $\focc<1$ means that the halo mass function is scaled down.
If \mmin\ decreases the total number of halos increases because less
massive halos are more abundant.
This requires smaller values of $\focc$ to keep the total
number of halos hosting LAEs equal to the number of LAEs in observations.

\begin{figure}
\includegraphics[width=0.47\textwidth]{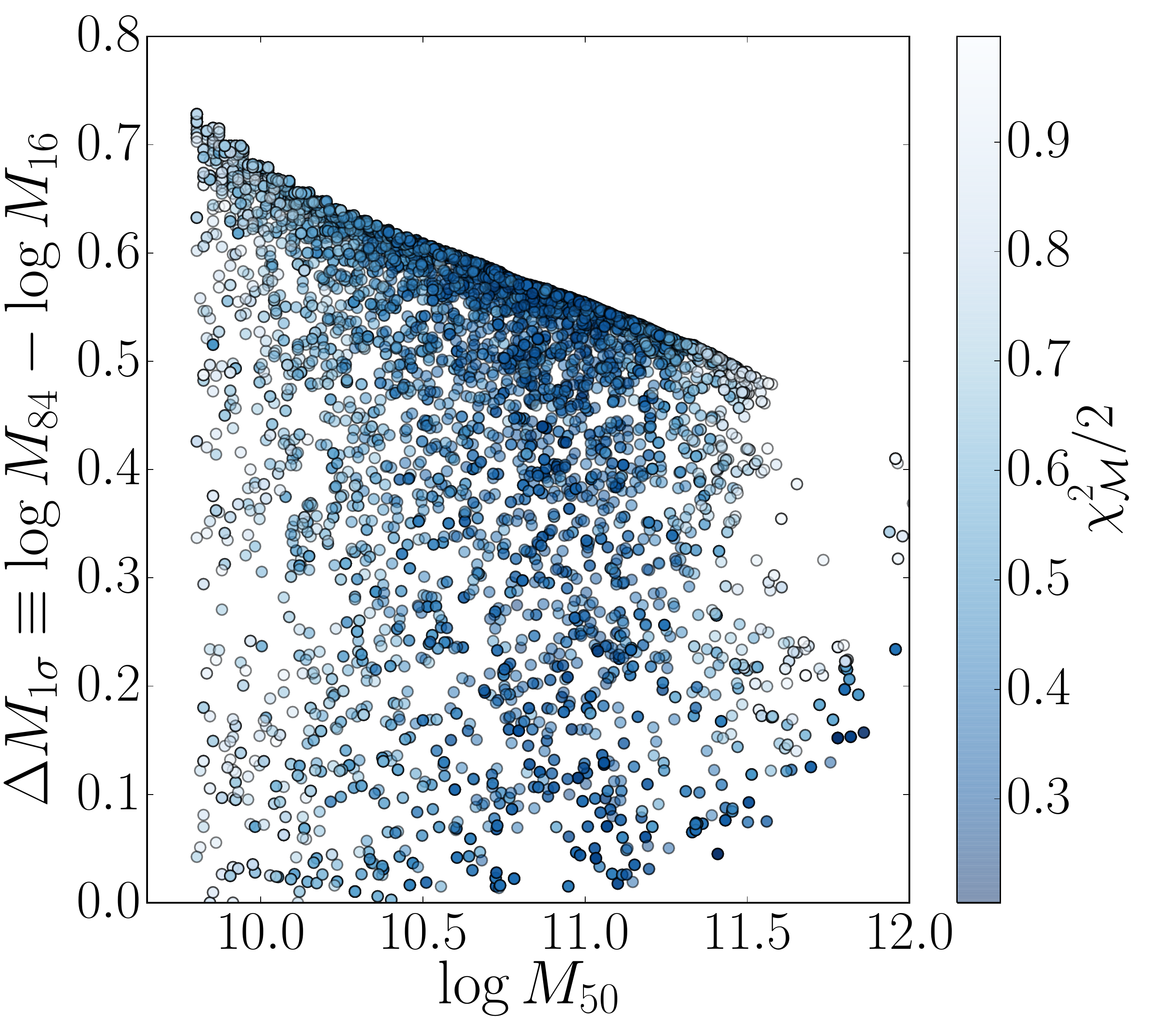}
\caption{Median mass, $\log M_{50}$, and mass width $\Delta \log
  M_{1\sigma}=\log M_{84} - \log M_{16}$, for all models with low
  $\chi^2_{\mathcal{M}}/2 < 1.0$. 
  The median mass spans two orders of magnitude with a mean value
  around $10^{10.5}$\hMsun.   
  The mass width $\Delta \log M_{1\sigma}$ has a median of $0.55$ dex (a factor
  of $\sim3$ in mass) with a maximum value of $0.7$ dex (a factor of
  5). This means that all models consistent with observations have a
  very narrow range in halo mass.}
\label{fig:mmed}
\end{figure}

\subsection{Median Halo Mass and Halo Mass Width within Models}

We now compute the median mass, $\log M_{50}$, and the $1\sigma$ halo mass
width, $\Delta \log M_{1\sigma} \equiv \log M_{84} - \log M_{16}$, of the
LAE hosting halos for each model  $\mathcal{M}$ (here $M_{p}$
represents the $p$ percentile of the mass distribution); $\log M_{50}$
and $M_{1\sigma}$ allow a direct comparison between our results and
previous results in the literature
\citep[e.g.][]{Hayashino2004,Gawiser2007,Ouchi2010,Bielby16} that used
a more simplified semi-analytical approach. 

In Fig. \ref{fig:mmed} we show the $\log M_{50}$-$\Delta \log M_{1\sigma}$
plane for the models selected to have $\chi^{2}_{\mathcal{M}}/2 < 1$,
which roughly correspond to the $1\sigma$ region in the posterior
distribution for the model parameters. The color encodes the
$\chi^{2}_{\mathcal{M}}/2$ value. The median mass has a wide distribution
spanning two orders of magnitude.

From this distribution we obtain $\log M_{50} = 10.8\pm 0.6$ or
equivalently $M_{50} = 6.3^{+18.8}_{-4.7}\times 10^{10}$\hMsun.  
The $1\sigma$ uncertainty for this median value is estimated from the
$16$ and $84$ percentile values in the $\log M_{50}$ distribution.  

Our result is consistent within the statistical uncertainties with
previous estimates reported by
 \citet{Bielby16} ($M_{50} = 10^{11.0\pm0.6} \hMsun$),
 \citet{Gawiser07} ($M_{50}= 10^{10.9\pm0.9} \hMsun$) and
 \citet{Ouchi2010} ($M_{50} = 10^{10.8^{+1.8}_{-0.8}} \hMsun$) 
using semi-analytical approaches.   Our result is slightly below the  \citet{Bielby16}
 median mass estimation. This could probably be attributed to the fact that  \citet{Bielby16}
only represent a particular field of the universe while our $\log M_{50}$  represents the
median mass of the universe after taking into account cosmic variance that allows lower mass 
halos to be also consistent with observations.

Figure \ref{fig:mmed} also shows something 
that semi-analytical approaches were not able to predict.
The mass width, $\Delta \log M_{1\sigma}$, that is the width of the mass
distribution for a given model with fixed $\mmin$ and $\mmax$, has a
median value and $1\sigma$ uncertainty of $\Delta \log M_{1\sigma} =
0.51^{+0.10}_{-0.30}$ dex.
This means that the mass range for halos hosting LAEs is very narrow.
There is only a factor of $2$ to $4$ between the lower and upper mass
boundary of the central 68 percentile of the mass distribution. 
We emphasize that \mmax\ can still be very large while the width is
small due to the asymmetry in the dark matter halo mass function.

Summarizing, the median mass could be anything in the range
$10^{10.2}$\hMsun\ and $10^{11.4}$\hMsun\ (a $2.0$ dex range), 
but the 1-$\sigma$ width of the mass distribution is highly constrained 
to be between $0.2$ and $0.6$ dex.

In Fig. \ref{fig:corr} we show the computed
$ACF_{\mathcal{M}}$ of models with $\lmmin=10.5$ and different values
of $\dlm$. We can see that the clustering gets slightly stronger for larger
values of \dlm. Nevertheless, due to the large impact of cosmic
variance at the volume of the current observations all the models  are
basically consistent within errors. The last result together with the
large Poissonian observational error in the ACF explain the current
difficulty to put tighter constrains in \lmmax\ in our model.

\subsection{Constraining Dark matter halos mass  with cosmic variance}

Fig. \ref{fig:cosmicv0}  shows the  halo number distribution (HND) in
the  mock fields of the simulation for different models
$\mathcal{M}$. 
By simple inspection one can infer that the distribution width
increases with \mmin.

\begin{figure}
\includegraphics[width=0.49\textwidth]{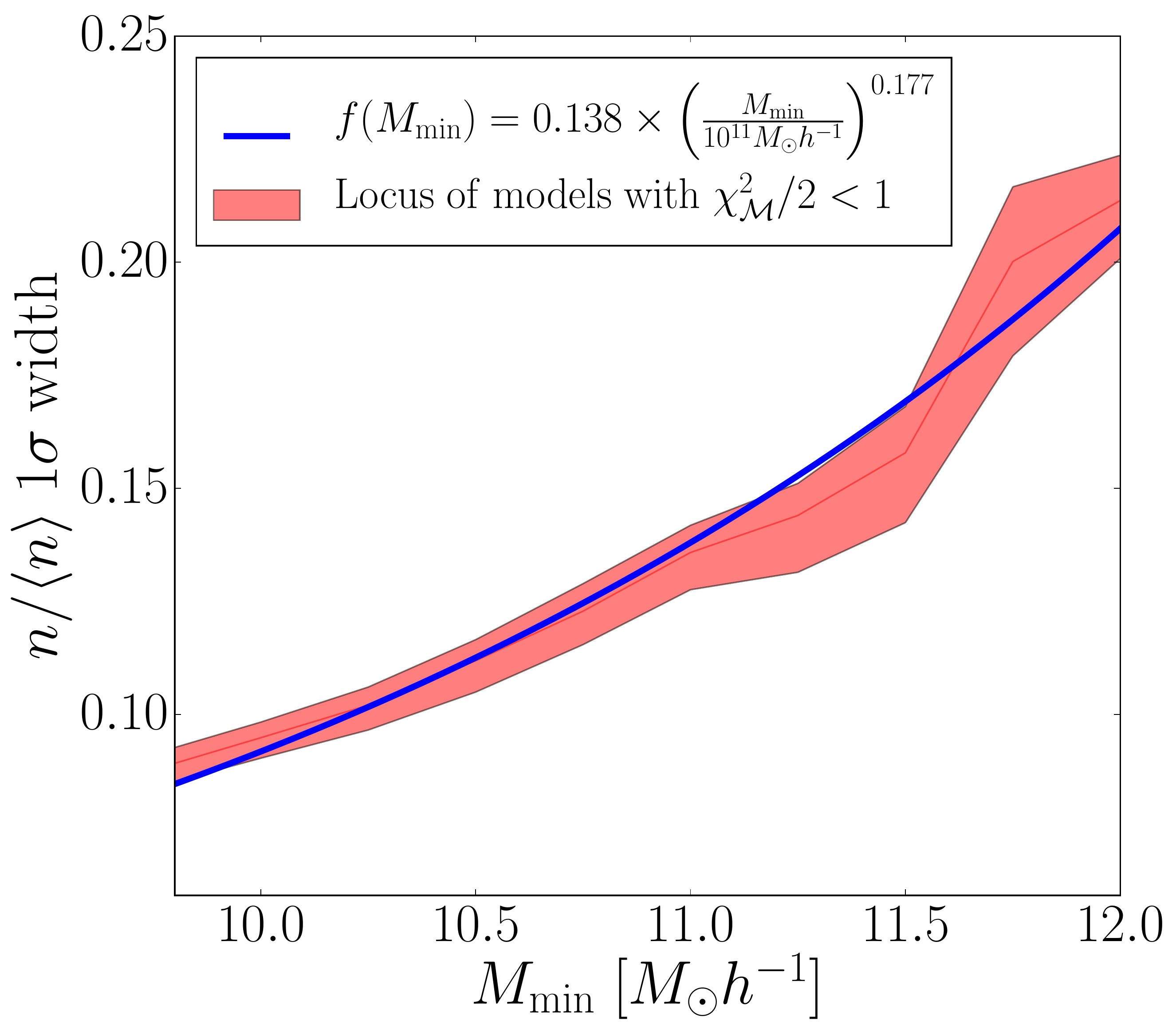}
\caption{$1\sigma$ width of the number density distribution over the
  as a function of \mmin. 
  Only the models that have a good match with observations
  ($\chi_{\mathcal M}/2 < 1$) are included.
  The strong dependence of $\langle
 n\rangle$ with \mmin was
  parameterized as a power law with the parameters shown in the legend.
\label{fig:cosmicv}}
\end{figure}

In Fig. \ref{fig:cosmicv} we confirm this trend by plotting the
HND $1\sigma$ width for good models ($\chi^{2}_{\mathcal{M}}/2 < 1$) as a function
of \lmmin. 
We find that considering all the $27$ mock fields the $1\sigma$ width, $W_{1\sigma}$
increases with \mmin following

\begin{equation}
W_{1\sigma} = (0.138\pm 0.002) \times \left( \frac{\mmin}{10^{11}\hMsun} \right)^{0.177\pm0.009}.
\label{eq:width}
\end{equation}

This result opens the possibility to constrain the \lmmin (as well as the
median mass) of LAEs by simply measuring the width of the distribution
of observed LAE along several observational fields. 
This idea has been already explored for $z>6$ galaxies
\citep{Robertson10}. 

To keep the validity of Eq. (\ref{eq:width}) there should at least be
$27$ fields (to follow the same numbers we use here) of size $\sim1{\rm
  deg^2}$ with the same observational conditions (filters, equivalent
width cuts). 
However, repeating the same kind of cosmic variance study for different
survey strategies should help to provide a constraint of the same
kind.

 To keep the validity of Eq. (\ref{eq:width}) there should be of the order of
 $27$ field or even a larger number. Each of these fields should have  $\sim1{\rm
  deg^2}$ in size and reproduce the same observational conditions of 
  \citet{Bielby16} in terms of filters, equivalent
width cuts .  However, repeating the same kind of cosmic variance 
study for different survey strategies should help to provide a equivalent
constraint. 

\subsection{Hints of larger uncertainties}

Fig. \ref{fig:like} shows an interesting feature for \mmin. 
The histogram (upper left) showing the probability for this parameter
given the observational constraints does not considerably decreases for lower
masses. 
This raises the following question. Down to which values of
\mmin\ does the probability significantly decreases? 

Lower values of \mmin\ would imply a lower \focc\ and larger
uncertainties on the parameters \mmin\, \focc and $M_{50}$.
In other words, the limitations in our priors for \mmin\ (which
set the limitations for the kind of numerical simulation we use) might
be underestimating the uncertainties and slightly biasing our results
for \mmin\, \focc\ and $M_{50}$.

Although Fig. \ref{fig:mmed} suggest that models with lower values of
$M_{\rm 50}$, i.e. lower values of \mmin, have a higher
$\chi^2_{\mathcal{M}}$ and could be thus discarded, we suggest that a
proper test of this hypothesis requires performing new cosmological
simulations to be able to probe \mmin\ masses below our limit of
$6\times  10^{9}$\hMsun. 

This highlights the main thesis of this paper. 
Namely, the high impact that cosmic variance can have on constraining
the paremeters of our model. 
With the current limitations in the cosmological simulations available
to the public we cannot significantly decrease or prior on \mmin.
We consider such test beyond the scope of this paper.

This extension to larger priors cannot change other central results of
our paper, such as the relationship between \focc\ and \mmin\, the
narrow mass range for individual models and the possibility to
constraint the halo mass with cosmic variance on the number density
counts. 
However, we suggest that future work building upon the
technique presented in this paper should count with simulations that
extend at least one order of magnitude below our limit $6\times
10^{9}$\hMsun.

%*************************************************************************

\section{Conclusions}

In this paper we studied the impact of cosmic variance and
observational uncertainties in constraining the mass range and
occupation fraction of dark matter halos hosting  LAEs.   
To this end we used the BolshoiP N-body simulation to construct  27 mock fields
with the same typical size  of observed fields at $z=3.1$ ($\sim 1
{\rm deg^2}$).      
In our model a dark matter halo with mass in the range $\mmin
<M_{h}<\mmax$ can only host one detectable LAE at most.  
We explored the parameter space determined by \mmin\ and \mmax\ using 
affine invariant Monte Carlo Markov-Chain minimization to match
 the observed  ACF and mean number density of LAEs.    
It is the first time that such a thorough exploration of the cosmic
variance impact on LAEs statistics is presented in the literature.

We find that once cosmic variance is taken into account, the
observational data can only put weak constraints on  \mmin\ and \focc. 
When we translate these loose constraints into the median mass, we find
a general consistency with previous works \citep{Hayashino2004,
  Gawiser07,Ouchi2008} specially with the most recent one
\citep{Bielby16}. 

This work also highlights the need to explore a wider range on the
prior of \mmin\ if one wants to better estimate the uncertainty on
\mmin, \focc\ and $M_{50}$.  
In this paper we use priors given by previous observations, but our
results suggest that once cosmic variance is explicitly taken into
account, even lower values of \mmin\ might be in fact consistent with
observations. 
We recommend that future work making use of the technique presented
in this paper should extend the prior for \mmin\ to masses at least
one order of magnitude below $6\times 10^{9}$\hMsun.  

Nevertheless, our analysis allowed us to draw two results that can
be used to put tighter constraints on \mmin\ and \focc\ once
upcoming large LAE surveys, such as the HETDEX project 
\citep{Hetdex2011} and the HSC ultra deep survey, are available: 

\begin{enumerate}
\item \focc\ is uniquely determined by \mmin, regardless of
  \mmax. A precise determination of \mmin\ will thus fix \focc. As a
  consequence of this, for a given  \mmin\ the median mass is also
  fixed.  

\item The width of the LAE number distribution function
  obtained over several fields of $\approx 1$ deg$^2$ is tightly
  correlated with \mmin. 
  That measurement with next generation surveys
  will be able to constrain \mmin within a factor of $\sim 2$.
\end{enumerate}

We also find that the 1-$\sigma$ width of the mass distribution is
highly constrained to be between $0.2$ and $0.6$ dex. 
This result can be used to test different models for LAEs formation
in a cosmological context to better understand why observable LAEs seem to be
constrained into a narrow halo mass range.

\section*{Acknowledgments} 

The authors acknowldge the referee, Nelson Padilla, for useful
remarks and questions that helped us to improve the quality and
clarity of our paper. The authors also acknowldge Valentino Gonzalez,
Eric Gawiser and Mark Dijkstra for the fruitful discussion and
suggestions that also contributed to improve the quality of this work. 

JEMR acknowledges ``CONICYT-PCHA/doctorado Nacional para
extranjeros/2013-63130316'' for their PhD scholarship support.  

JEFR acknowledges financial support from Vicerrector\'ia de
Investigaciones at Uniandes from a a FAPA project.

The authors gratefully acknowledge the Gauss Centre for Supercomputing
e.V. (\url{www.gauss-centre.eu}) and the Partnership for Advanced
Supercomputing in Europe (PRACE, \url{www.prace-ri.eu}) for funding the
MultiDark simulation project by providing computing time on the GCS
Supercomputer SuperMUC at Leibniz Supercomputing Centre (LRZ,
\url{www.lrz.de}). The Bolshoi simulations have been performed within the
Bolshoi project of the University of California High-Performance
AstroComputing Center (UC-HiPACC) and were run at the NASA Ames
Research Center.

\bibliographystyle{apj}
%\bibliography{references.bib}

\end{document}